\documentclass[conference]{IEEEtran}
\IEEEoverridecommandlockouts
\usepackage{cite}
\usepackage{amsmath,amssymb,amsfonts}
\usepackage{algorithmic}
\usepackage{graphicx}
\usepackage{textcomp}
\usepackage{xcolor}
\usepackage{tabularx}
\usepackage{threeparttable}
\usepackage{booktabs} 
\usepackage{lineno}

\def\BibTeX{{\rm B\kern-.05em{\sc i\kern-.025em b}\kern-.08em
    T\kern-.1667em\lower.7ex\hbox{E}\kern-.125emX}}
\begin{document}

\title{A Deep Learning Framework for Nuclear Segmentation and Classification in Histopathological Images \\
}

\author{\IEEEauthorblockN{1\textsuperscript{st} Sen Yang}
\IEEEauthorblockA{\textit{College of Biomedical Engineering} \\
\textit{Sichuan University}\\
Chengdu, China \\
ys810137152@gmail.com}
\and
\IEEEauthorblockN{2\textsuperscript{nd} Jinxi Xiang}
\IEEEauthorblockA{\textit{Department of Precision Instruments} \\
\textit{Tsinghua University}\\
Beijing, China \\
derek.hsiang17@gmail.com}
\and
\IEEEauthorblockN{3\textsuperscript{rd} Xiyue Wang}
\IEEEauthorblockA{\textit{College of Computer Science} \\
\textit{Sichuan University}\\
Chengdu, China \\
xiyue.wang.scu@gmail.com}
}

\maketitle

\begin{abstract}
Nucleus segmentation and classification are the prerequisites in the workflow of digital pathology processing. However, it is very challenging due to its high-level heterogeneity and wide variations. This work proposes a deep neural network to simultaneously achieve nuclear classification and segmentation, which is designed using a unified framework with three different branches, including segmentation, HoVer mapping, and classification. 
The segmentation branch aims to generate the boundaries of each nucleus. The HoVer branch calculates the horizontal and vertical distances of nuclear pixels to their centres of mass.    	
The nuclear classification branch is used to distinguish the class of pixels inside the nucleus obtained from segmentation.

\end{abstract}


\section{Methods}
\begin{figure*}[h!]
	\centerline{\includegraphics[width=15cm]{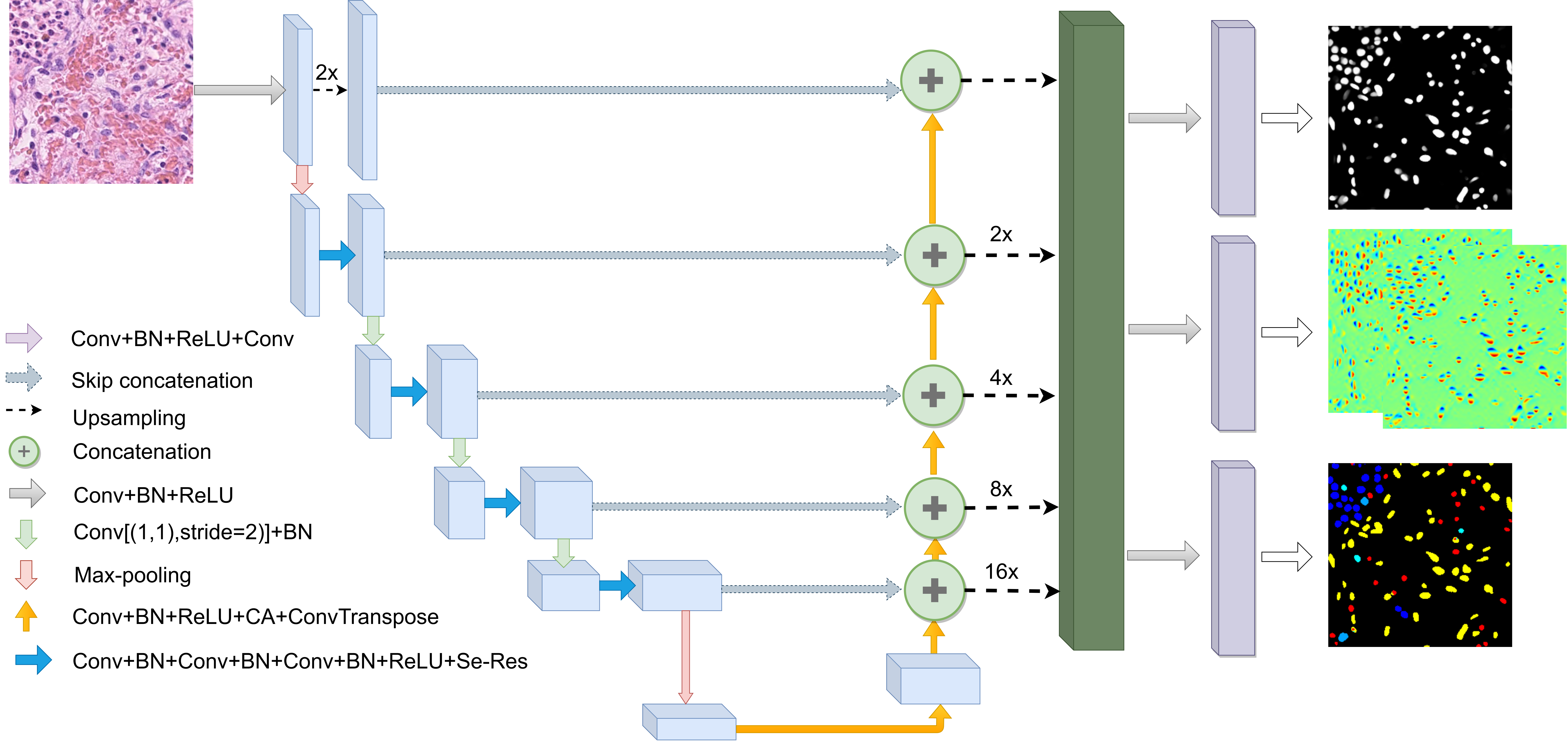}}
	\caption{Pipeline of our proposed nuclear instance segmentation and classification algorithm. 
	} 
\label{main}
\end{figure*}
We propose a unified model to achieve the simultaneous nuclear instance segmentation and classification on histopathological images, which is illustrated in Fig.~\ref{main}. Benefiting from the design of HoVer-Net, we employ a three-branch structure for simultaneous nuclear segmentation, HoVer map production, and pixel-level nuclear classification. The segmentation branch aims to depict the boundary of each nucleus. The HoVer branch generates horizontal and vertical maps by calculating horizontal and vertical distances of nuclear pixels to their centres of mass, which is used in post-processing to achieve precise nuclear segmentation. The nuclear classification branch divides the nucleus into six categories: epithelial, lymphocyte, plasma, eosinophil, neutrophil, or connective tissue. 

The three branches have shared encoder and decoder but different heads. Our network follows the general encoder-decoder architecture of the popular U-Net model \cite{ronneberger2015u}. But we replace the convolution blocks in the encoder part of the original U-Net with a more elaborate SE-Res module originally proposed in \cite{hu2018squeeze}. The SE-Res module employs a gating mechanism with the sigmoid activation to capture channel-wise feature dependencies, which helps to enhance more informative features and suppress less useful ones. In the decoder part of the network, we embed the CA module \cite{hou2021coordinate} at each resolution level in order to capture cross-channel, direction-aware, and position-sensitive information. The skip connections of the original U-Net are kept between the encoder and decoder blocks, which helps aggregate features of inter-channel relationships and precise positional information at the decoder to get more accurate segmentation results.

The prediction heads in the three branches have similar structures but different weights, which are composed of convolution, batch normalization (BN), and ReLU. The loss functions adopt the Dice + cross-entropy (CE) and mean squared error (MSE) in the segmentation branch and HoVer branches, respectively. Considering the class imbalance of  different types of nuclei, the loss function of the classification branch employs  the Dice and weight CE. These loss functions are defined as follows.

\begin{equation}
L_{\text {seg}}=L_{CE}+L_{\text {Dice}}
\end{equation}
\begin{equation}
L_{\text {HoVer}}=L_{MSE}(y, \hat{y})=\|\widehat{y}-y\|_{2}^{2}
\end{equation}
\begin{equation}
L_{\text {cls}}=L_{WCE}+L_{\text {Dice}}
\end{equation}

\begin{equation}
L_{\text {WCE}}(y, \hat{y})=-\frac{1}{N} \sum_{i=1}^{N} \sum_{k=1}^{C} w_{k} y_{i, k} \log \hat{y}_{i, k}
\end{equation}

\begin{equation}
L_{\text {Dice}}=-\frac{2}{|C|} \sum_{k \in C} \frac{\sum_{i} y_{i, k} \hat{y}_{i, k}+\varepsilon}{\sum_{i} y_{i, k}+\sum_{i} \hat{y}_{i, k}+\varepsilon}
\end{equation}
\noindent where $y$ and $ \hat{y}$ represent the ground truth label and the predicted label, respectively. The $w_k$ denotes the class weights, which are set empirically to 2, 2, 3, 4, 4, 2, and 1 for the epithelial, lymphocyte, plasma, eosinophil, neutrophil, connective tissue, and background classes, respectively. When the $w_k$ equals 1,  the $L_{\text {WCE}}$ becomes the traditional $L_{\text {CE}}$.

After the network training, our post-processing follows the method used by HoVer-Net \cite{graham2019hover}, which helps drive more accurate classification and segmentation results.

\section{Results}
\subsection{Experimental setups}
Standard real-time data augmentation methods such as horizontal flipping, vertical flipping, random rescaling, random cropping, and random rotation are performed to make the model invariant to geometric perturbations. Moreover, RandomHSV is also adopted to randomly change the hue, saturation, and value of images in the hue saturation value (HSV) color space, making the model robust to color perturbations. The Adam optimizer \cite{kingma2014adam} is used as the optimization method for model training. The initial learning rate is set to 0.0003, and reduced by a factor of 10 at the 25th and the 35th epoch, with a total of 50 training epochs. The min-batch size is set as 32. All models are implemented using the PyTorch framework and all experiments are performed on a workstation equipped with an Intel(R) Xeon(R) E5-2680 v4 2.40GHz CPU and four 32 GB memory NVIDIA Tesla V100 GPU cards.

\subsection{Experimental results}
Two evaluation metrics are used for algorithm validation, including multi-class panoptic quality ($mPQ^+$) and  multi-class coefficient of determination ($R^2$), which keep the same with the organizer of the challenge \cite{graham2021conic}. Table~\ref{tab} shows our results by comparing them with other methods. 

\begin{table}[htbp]
	\centering
	\caption{Comparison between our results and other methods}
	\begin{tabular}{lcc}
		\toprule
		Methods & $mPQ^+$   & $R^2$ \\
		\midrule
		HoVer-Net (baseline) & 0.29558  & -0.42802  \\
		MaskRCNN & 0.35460  & -0.19821  \\
		Ours  & 0.43419  & 0.56484  \\
		\bottomrule
	\end{tabular}%
	\label{tab}%
\end{table}%

As shown in  Table~\ref{tab}, it is seen that our method outperforms previous HoVer-Net and MaskRCNN by about 0.14 and 0.8 in terms of $mPQ^+$ metric. The results demonstrate the effectiveness of our proposed nuclear classification and segmentation algorithm. 

\bibliographystyle{ieeetran}
\bibliography{references}
\end{document}